\documentstyle[epsfig]{mn}
\begin{document}

\title
[The Ursa Major Cluster of Galaxies]
{The Ursa Major Cluster of Galaxies.~III.~Optical observations of dwarf
galaxies and the luminosity function down to $M_R=-11$ }

\author[Neil Trentham, R.~Brent Tully and Marc A.~W.~Verheijen ]
{
Neil Trentham$^{1}$, R.~Brent Tully$^{2}$ and
Marc A.~W.~Verheijen$^{3}$ \\
$^1$ Institute of Astronomy, Madingley Road, Cambridge, CB3 0HA.\\
$^2$ Institute for Astronomy, University of Hawaii,
2680 Woodlawn Drive, Honolulu HI 96822, U.~S.~A.\\
$^3$ NRAO-Array Operations Center, P.~O.~Box 0, Socorro NM 87801, U.~S.~A.\\
}
\maketitle

\begin{abstract} 
{Results are presented of a deep optical survey of the Ursa Major Cluster,
a spiral-rich cluster of galaxies at a distance of
18.6 Mpc which contains about 30\% of the light but only 5\% of the
mass of the nearby Virgo Cluster.
Fields around known 
cluster members and a pattern of
blind fields along the major and minor axes of the cluster were studied
with mosaic CCD cameras on the Canada-France-Hawaii Telescope.
The dynamical crossing time for the Ursa Major Cluster is only slightly less 
than a Hubble time.
Most galaxies in the local Universe exist in similar moderate density 
environments.
The Ursa Major Cluster is therefore a
good place to study the statistical properties of dwarf galaxies 
since this structure is at an evolutionary stage 
representative of typical environments yet has enough galaxies that
reasonable counting statistics can be accumulated.  The main observational
results of our survey are: 
\vskip 1pt
\noindent (i) The galaxy luminosity function is flat, with a
logarithmic slope $\alpha = -1.1$ for $-17 < M_R < -11$
from a power-law fit.
The error in $\alpha$ is likely to be less than 0.2 and is dominated by 
systematic errors, primarily associated with uncertainties in
assigning membership to specific galaxies. 
This faint end slope is quite different to what was seen in the
Virgo Cluster, where $\alpha = -2.26 \pm 0.14$.
\vskip 1pt
\noindent (ii) Dwarf galaxies are as frequently found to be blue dwarf 
irregulars as red dwarf spheroidals in the blind cluster fields.  The density
of red dwarfs is significantly higher in the
fields around luminous members than in the blind fields. 
\vskip 1pt
The most important result is the {\it failure} to detect many dwarfs.  If the
steep luminosity function claimed for the Virgo Cluster were valid for Ursa 
Major then in our blind fields we should have found $\sim 10^3$ galaxies with
$-17<M_R<-11$ where we have found two dozen.  There is a clear deficiency of
dwarfs compared with the expectations of hierarchical clustering theory.
It is speculated that the critical difference between the Virgo and Ursa Major
clusters is the very different dynamical collapse times, which probably
straddle the timescale for reionization of the Universe.  Dwarf galaxies in 
the proto-Virgo environment probably formed before the epoch of reionization.
The equivalent dwarf halos in the proto-Ursa Major environment probably only
formed after the epoch of reionization, when the conditions for star 
formation were inhospitable.
}
\end{abstract}  
\vskip 40pt


\section{Introduction}

The galaxy luminosity function $\phi (L)$, 
defined as the number density of
galaxies per unit luminosity $L$, is an important probe of
the physical processes that leads to galaxy formation. 
The luminosity function depends on both the primordial fluctuation
spectrum and on the physics governing star formation.
The luminosity function has been a popular diagnostic since it is so
straightforward to measure, at least at high luminosities, and it is
expected to 
tell us something about a more fundamental parameter, the mass
function.
Early theoretical attempts  were successful at
reproducing the general form of the luminosity function 
(e.g.~White \& Rees 1978),
which decreases
monotonically with increasing luminosity, and decreases very steeply indeed 
above
some characteristic luminosity   
$L_* \sim 2\times10^{10}~h_{75}^{-2} {\rm L}_{B\, \odot}$ 
(Schechter 1976; where $h_{75} = {\rm H}_{\circ}/(75 
\,{\rm km}\,{\rm s}^{-1}\,{\rm Mpc}^{-1})$; 
H$_{\circ}$ is the
Hubble Constant).
More recent theoretical models 
(e.g.~Baugh et al.~1998, Somerville \& Primack 1999, 
Diaferio et al.~1999) have been based on a semi-analytic
approach to the study of galaxy formation and are able to make more detailed
predictions.  

The luminosity function of galaxies in the field is normally determined
based on an imaging survey over a chosen angular region of the
sky with follow-up redshift measurements of the galaxies (e.g.~the
Las Campanas Redshift Survey of Lin et al.~1996 and the Autofib survey of  
Ellis et al.~1996).   From Hubble's law, one
has distances and hence can compute luminosities and construct
a luminosity function for a cone volume with the observer at the origin.
The luminosity functions measured this way tend to be very well-determined
at the bright end but poorly-determined at the faint end.
Most galaxies in a magnitude-limited sample are distant
luminous galaxies, not nearby low-luminosity galaxies.
The clumpiness of the distribution of galaxies and the fact that the
faintest galaxies are drawn from a very small region means the
normalization of the faint end relative to the brighter end is usually dubious.
Imaging to extremely deep limits does not help
since the angular coverage on the sky is then
necessarily small -- for
example there are only two galaxies in the faintest bin of the Keck
survey of Cowie et al.~(1996).  
Galaxies in the Local Group are known down to very faint limits ($M_V \sim
-8.5$), but the local luminosity function (van den Bergh 1992, 2000) 
suffers from poor counting statistics at all luminosities, in addition to 
possible
incompleteness at the very faint end.  For example, the Cetus dwarf
(Whiting et al.~1999), with
$M_V \sim -10$, was discovered only last year.   

An alternative approach to measuring the luminosity function down to very
low luminosities is as follows.  
The vast majority of low-luminosity galaxies (the ``dwarf galaxies'') have low
surface-brightnesses and follow (albeit with some scatter) the
absolute magnitude vs.~central
surface-brightness correlation shown in Figure 1 of Binggeli (1994).
Red dwarf spheroidal (dSph) and blue dwarf irregular (dIrr) galaxies both
have azimuthally-averaged light profiles that are exponential, and
follow the same
absolute magnitude vs.~central surface-brightness
correlation (Binggeli \& Cameron 1991, Binggeli 1994).  
Therefore if one finds a low surface-brightness galaxy of a given apparent
magnitude in a cluster, it is far more likely to be a 
low-luminosity member of that cluster than a
high-luminosity background
galaxy.  The converse is true if one finds a high surface-brightness galaxy of
the same apparent magnitude.  These statements can be made more quantitative
by observations of blank sky fields -- there is a marked absence of low
surface-brightness galaxies in these fields, as we shall see later in this
paper.  
One place where this kind of technique has been employed very successfully
is in the Virgo Cluster, where Phillipps et al.~(1998) find a very steep
galaxy luminosity function, with $\alpha \sim -2$ (here $\alpha$ is
the logarithmic slope of the luminosity function: $\phi (L) \sim L^{\alpha}$)
between $M_R = -14$ 
and $M_R = -11$, close to the predicted slope of the galaxy mass function
from Press \& Schechter (1974) theory, assuming the cold dark matter
fluctuation spectrum of Bardeen et al.~(1986). 
These results probe significantly deeper than the well-known Virgo luminosity
function of Sandage, Binggeli \& Tammann
(1985; see also Impey, Bothun \& Malin 1987),
or the studies of nearby groups of Tully (1988) or  
of Ferguson \& Sandage (1991), all of whom found far shallower luminosity
functions.  
The Virgo results, however, do not necessarily tell us anything about the
field luminosity function, which is what is important for cosmology.
That is because the Virgo Cluster is a dense environment where the
galaxies formed early and where
galaxy-galaxy interactions probably played an anomalously important role
in shaping present-day galaxy properties.

So we now apply these techniques to the Ursa Major Cluster, a  
large but 
diffuse cluster of spiral galaxies at distance of 18.6 Mpc (Tully \&
Pierce 2000), close to the
Virgo Cluster, and attempt to determine the luminosity function of
low-luminosity galaxies there.  
This cluster is very loosely held together and has no appreciable X-ray
halo.  It is quite different from clusters of elliptical galaxies
like Virgo (itself not a particularly rich cluster). 
The velocity dispersion of the Ursa Major Cluster is low (148 km s$^{-1}$,
compared to 715 km s$^{-1}$ for Virgo), and its mass is about 1/20 of
the mass of the Virgo Cluster (Tully~1987$b$).   
The results for the Ursa Major Cluster are expected to be be far
more representative of what the field galaxy luminosity function will look
like at the faint end because most galaxies in the Universe
exist in diffuse spiral-rich environments.

A deep optical survey of sufficient angular area for good
Poisson statistics is made possible
in nearby clusters like Ursa Major 
by the advent of mosaic CCDs on large telescopes.  In this work we
used the UH8K (Metzger, Luppino \& Miyazaki 1995) and 
CFH12K (Cuillandre et al.~1999) mosaic cameras 
on the Canada-France-Hawaii Telescope (hereafter CFHT)
to image the cluster along its major
and minor axes, along with a number of pointed observations around
known cluster members.
Additionally we have obtained
HI images with the Very Large Array (hereafter
VLA) of the fields along the major and minor axes. 
In this paper we present all the optical data.  The radio data
are presented elsewhere (Verheijen et al.~2001).  
The fields around known cluster members already have HI data
available, from the Westerbork Synthesis Radio Telescope (hereafter WSRT;
Verheijen 1998). 
We also have taken pointed multicolour images of probable
new members and present those data here. 

Additionally, we will be able to use the results to determine the 
colour distribution of low-luminosity galaxies in the cluster, which
constrains star formation
histories, and give attention to the morphology-density
relation of dwarf galaxies in the Ursa Major Cluster. 
It should be clear from our data whether the dwarf galaxies congregate
around the giant galaxies or are more uniformly distributed within the
cluster, and whether or not
the answer depends on the dwarf galaxy morphology, HI mass,
and colour.  In the local Universe, these three properties certainly play
an important role: red gas-poor dwarf spheroidals cluster
around giant galaxies, whereas blue gas-rich dwarf irregulars are less
correlated in position with giant galaxies
(Binggeli, Tarenghi \& Sandage 1990).   

This paper is organized as follows.  In Section 2 we describe the observations
and basic data processing.  In Section 3 we describe how we perform
photometry and outline how we identify
plausible cluster candidate members.  In Section 4
we look at the sample so constructed in more detail and describe how
confident we are about each identification.  Using this sample, we determine
luminosity functions, colour distributions, and morphology-density relations
in Sections 5 -- 7.  Finally in Section 8 we summarize and
attempt to put all the results
together to obtain a coherent picture of the Ursa Major Cluster. 
 
\section{Observations and Data Reduction} 

\subsection{General strategy} 

Our basic observing strategy for this project was as follows: 
in March 1996
and March 1999 we observed regions in the Ursa Major Cluster using
large-format mosaic CCDs with the
intention of finding low luminosity galaxies, which we identified based on
their low surface brightnesses.  A number of background fields were taken
for comparison purposes.  Pointed three filter observations of the 
candidates
that we identified were then taken in February 2000
using a single-chip CCD so that colours could be measured.
In 1996 we mostly observed fields centered
on known cluster members (see Figure 1 and Table 1).  
In 1999 we observed contiguous fields along the major and
minor axes of the cluster (again see Figure 1 and Table 1; the field 
designations
we used are presented there).   
All the fields studied in 1999 were also observed with
the HI line receiver at the
VLA and all but the `blank' fields observed in 1996 were observed with the
HI line receiver at the WSRT.  A number of dwarf candidates turned out to be 
HI gas-rich and we are 
therefore able to confirm that they really are members from their observed
velocities. 

\begin{figure}
\begin{center} 
\vskip-2mm
\epsfig{file=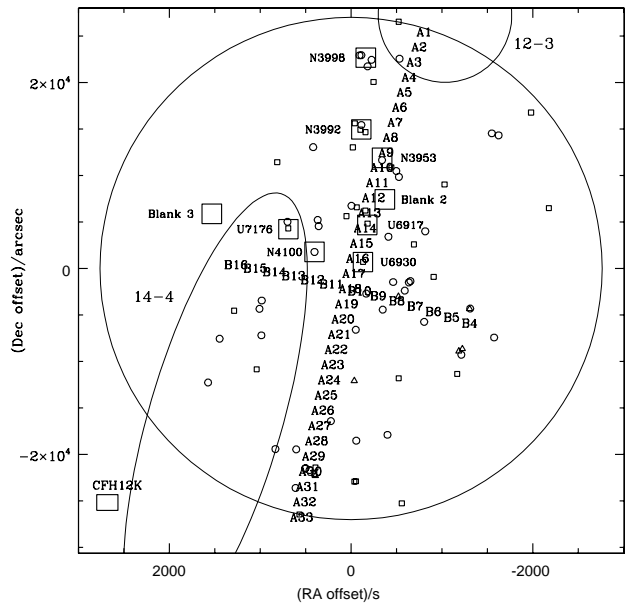, width=8.65cm}
\end{center}
\vskip-4mm
\caption{
The field positions, shown
as offsets from the cluster center which we take to be
$(\alpha(2000), \delta (2000)) = (11^{\rm h} 59^{\rm m} 28.3^{\rm s},
49^{\circ} 05^{\prime} 18^{\prime \prime})$. 
The large squares represent the UH8K fields observed in 1996 (see Table 1).  
The letters/numbers represent the CFH12K fields observed in 1999; the
designations here are as in Table 1.  The size of the CFH12K field is
shown in the rectangle at the bottom left.  The 
small open circles represent known
NGC members.  The small open squares represent known UGC members.  The
small open triangles represent other known members.
The large circle represents the approximate extent of the Ursa Major Cluster;
i.e.~the region within 7.5 degrees of the cluster center as defined
above.  
The positions of the nearby 12$-$3 and 14$-$4 Groups
(Tully 1987$a$) 
are also shown.
}
\end{figure}

For the regions of the sky we study, Galactic extinction is  
small: $E(B-V) < 0.05$ mag 
but varies slightly from field to field and we use the measurements of
Schlegel, Finkbeiner
\& Davis 1998 to correct our data for this effect.  

\begin{table*}
\caption{Fields observed} 
{\vskip 0.75mm}
{$$\vbox{
\halign {#\hfil \hfil && \quad #\hfil \hfil \cr
Field &  
$\alpha$ (2000) & $\delta$ (2000) & Known Members ($R$ mag)$^{*}$
 & Other galaxies ($R$ mag)$^{\dagger}$  
&\cr 
\noalign{\smallskip} \noalign{\smallskip}
\noalign{\smallskip}
{\bf 1996}  &   &     &     & &\cr
Blank 1 &
$11 \, 37 \, 32.4 \, $&
$56 \, 55 \, 14 \, $&
  &  &\cr
Blank 2 &
$11 \, 52 \, 38.9 \, $&
$51 \, 09 \, 03 \, $&
  &  &\cr 
NGC 3953 field &
$11 \, 53 \, 48.6 \, $&
$52 \, 23 \, 36 \, $&
NGC 3953 (9.66) &  &\cr
UGC 6917 field &
$11 \, 56 \, 30.4 \, $&
$50 \, 23 \, 16 \, $&
UGC 6917 (12.16) &  &\cr
NGC 3998 field &
$11 \, 56 \, 47.6 \, $&
$55 \, 23 \, 12 \, $&
NGC 3998 (9.55) & NGC 3977 (13.4)$^{*0}$ &\cr
 &  &    & NGC 3972 (11.90) & &\cr
 &  &    & NGC 3990 (12.08) & &\cr
UGC 6930 field &
$11 \, 57 \, 19.1 \, $&
$49 \, 16 \, 55 \, $&
UGC 6930 (11.71) &  &\cr
NGC 3992 field &
$11 \, 57 \, 36.2 \, $&
$53 \, 14 \, 29 \, $&
NGC 3992 (9.55) &   &\cr
 &  &    & UGC 6923 (12.97) & &\cr
 &  &    & UGC 6969 (14.32) & &\cr
 &  &    & UGC 6940 (15.65) & &\cr
NGC 4100 field &
$12 \, 06 \, 08.1 \, $&
$49 \, 34 \, 59 \, $&
NGC 4100 (10.62) &  &\cr
UGC 7176 field &
$12 \, 10 \, 55.3 \, $&
$50 \, 15 \, 49 \, $&
UGC 7176 (15.61) &  &\cr
Blank 3 &
$12 \, 24 \, 52.3 \, $&
$50 \, 43 \, 23 \, $&
  &  &\cr
           &   &     &     & &\cr
{\bf 1999} &   &     &     & &\cr  
A01 & $11 \, 46 \, 09.2 \, $ & 
$56 \, 09 \, 07 \, $ & 
                &  NGC 3888 (11.8)$^{*1}$   &\cr 
 & & & & NGC 3898 ($B$ = 11.6)$^{*2}$  &\cr
A02 & $11 \, 47 \, 05.4 \, $ & 
$55 \, 42 \, 15 \, $ &
                 &  NGC 3850 (13.2)$^{*3}$  &\cr  
A03 & $11 \, 48 \, 00.2 \, $ & 
$55 \, 15 \, 20 \, $ &
                 &                &\cr  
A04 & $11 \, 48 \, 53.9 \, $ & 
$54 \, 48 \, 25 \, $ &
                 &                &\cr  
A05 & $11 \, 49 \, 46.3 \, $ & 
$54 \, 21 \, 28 \, $ &
                 &                &\cr  
A06 & $11 \, 50 \, 37.6 \, $ & 
$53 \, 54 \, 30 \, $ &
                 &                &\cr  
A07 & $11 \, 51 \, 27.8 \, $ & 
$53 \, 27 \, 30 \, $ &
                 & UGC 6828 (13.5)$^{*4}$ &\cr  
A08 & $11 \, 52 \, 17.0 \, $ & 
$53 \, 00 \, 29 \, $ &
                 &                &\cr  
A09 & $11 \, 53 \, 05.2 \, $ & 
$52 \, 33 \, 26 \, $ &
NGC 3953 (9.66)  &                &\cr  
A10 & $11 \, 53 \, 52.3 \, $ & 
$52 \, 06 \, 23 \, $ &
UGC 6840 (13.35)  &                &\cr  
A11 & $11 \, 54 \, 38.6 \, $ & 
$51 \, 39 \, 19 \, $ &
                 &                &\cr  
A12 & $11 \, 55 \, 23.9 \, $ & 
$51 \, 12 \, 13 \, $ &
                 &                &\cr  
A13 & $11 \, 56 \, 08.3 \, $ & 
$50 \, 45 \, 06 \, $ &
UGC 6922 (13.65)  &                &\cr  
 & & & UGC 6956 (13.83)$^{*5}$ & &\cr
A14 & $11 \, 56 \, 51.9 \, $ & 
$50 \, 17 \, 59 \, $ &
UGC 6917 (12.16)    &                &\cr  
A15 & $11 \, 57 \, 34.6 \, $ & 
$49 \, 50 \, 50 \, $ &
                 &                &\cr  
A16 & $11 \, 58 \, 16.6 \, $ & 
$49 \, 23 \, 40 \, $ &
UGC 6930 (11.71)   &                &\cr  
A17 & $11 \, 58 \, 57.8 \, $ & 
$48 \, 56 \, 30 \, $ & 
                &                &\cr  
A18 & $11 \, 59 \, 38.3 \, $ & 
$48 \, 29 \, 18 \, $ & 
                &                &\cr  
A19 & $12 \, 00 \, 18.0 \, $ & 
$48 \, 02 \, 06 \, $ & 
                &                &\cr 
A20 & $12 \, 00 \, 57.0 \, $ & 
$47 \, 34 \, 52 \, $ & 
                & PC1200+4755 (16.0)$^{*6}$  &\cr  
A21 & $12 \, 01 \, 35.4 \, $ & 
$47 \, 07 \, 38 \, $ & 
                &                &\cr  
A22 & $12 \, 02 \, 13.1 \, $ & 
$46 \, 40 \, 23 \, $ & 
                &                &\cr  
A23 & $12 \, 02 \, 50.2 \, $ & 
$46 \, 13 \, 08 \, $ & 
                &                &\cr  
A24 & $12 \, 03 \, 26.8 \, $ & 
$45 \, 45 \, 51 \, $ & 
                &                &\cr  
A25 & $12 \, 04 \, 02.7 \, $ &
$45 \, 18 \, 34 \, $ &
                &                &\cr
A26 & $12 \, 04 \, 37.9 \, $ &
$44 \, 51 \, 16 \, $ &
                &                &\cr
A27 & $12 \, 05 \, 12.8 \, $ &
$44 \, 23 \, 58 \, $ &
NGC 4051 (9.88)  &                &\cr
A28 & $12 \, 05 \, 47.0 \, $ &
$43 \, 56 \, 39 \, $ &
                &                &\cr
A29 & $12 \, 06 \, 20.7 \, $ &
$43 \, 29 \, 19 \, $ &
                &                &\cr
A30 & $12 \, 06 \, 54.0 \, $ &
$43 \, 01 \, 58 \, $ &
             NGC 4111 (9.95) & CGCG215$-$022 (12.8)$^{*7}$      &\cr
  &  & &     NGC 4117 (12.47) & UGC 7069 (14.4)$^{*8}$      &\cr
  &  & &     NGC 4118 (14.82) &    &\cr
  &  & &     UGC 7094 (13.70) &    &\cr
  &  & &     UGC 7089 (12.77) &    &\cr
  &  & &     1203+43 (15.79) &    &\cr
A31 & $12 \, 07 \, 26.7 \, $ & 
$42 \, 34 \, 37 \, $ & 
NGC 4143 (10.55)   &                &\cr  
A32 & $12 \, 07 \, 58.9 \, $ & 
$42 \, 07 \, 16 \, $ & 
                &                &\cr  
A33 & $12 \, 08 \, 30.7 \, $ & 
$41 \, 39 \, 54 \, $ & 
                &                &\cr 
\noalign{\smallskip }
\noalign{\smallskip}\cr}}$$}
\end{table*}

\begin{table*}
{\vskip 0.75mm}
{$$\vbox{
\halign {#\hfil \hfil && \quad #\hfil \hfil \cr
Field &
$\alpha$ (2000) & $\delta$ (2000) & Known Members ($R$ mag)$^{*}$
 & Other galaxies ($R$ mag)$^{\dagger}$
&\cr
\noalign{\smallskip} \noalign{\smallskip}
\noalign{\smallskip}
B04 & $11 \, 37 \, 49.8 \, $ &
$47 \, 27 \, 29 \, $ &
                 &                &\cr
B05 & $11 \, 41 \, 07.6 \, $ &
$47 \, 38 \, 00 \, $ &
         &   NGC 3811 (12.1)$^{*9}$  &\cr
B06 & $11 \, 44 \, 26.8 \, $ &
$47 \, 48 \, 08 \, $ &
                 &                &\cr
B07 & $11 \, 47 \, 47.2 \, $ &
$47 \, 57 \, 55 \, $ &
                 &                &\cr
B08 & $11 \, 51 \, 08.8 \, $ &
$48 \, 07 \, 22 \, $ &
1148+48 (16.12)   &                &\cr
B09 & $11 \, 54 \, 31.8 \, $ &
$48 \, 16 \, 25 \, $ &
                 &                &\cr
B10 & $11 \, 57 \, 55.8 \, $ &
$48 \, 25 \, 06 \, $ &
NGC 3985 (12.26)   &                &\cr
B11 & $12 \, 03 \, 04.0 \, $ &
$48 \, 37 \, 25 \, $ &
                 &  NGC 4047 (11.9)$^{*10}$ &\cr
B12 & $12 \, 06 \, 30.9 \, $ &
$48 \, 45 \, 08 \, $ &
                 &                &\cr
B13 & $12 \, 09 \, 58.7 \, $ &
$48 \, 52 \, 29 \, $ &
                 &                &\cr
B14 & $12 \, 13 \, 27.6 \, $ &
$48 \, 59 \, 26 \, $ &
                 &                &\cr
B15 & $12 \, 16 \, 57.4 \, $ &
$49 \, 06 \, 00 \, $ &
                 &                &\cr
B16 & $12 \, 20 \, 28.1 \, $ &
$49 \, 12 \, 09 \, $ &
                 &  UGC 7358 (13.2)$^{*11}$ &\cr
\noalign{\smallskip }
\noalign{\smallskip}\cr}}$$}
\begin{list}{}{}
\item[$^{*}$] Magnitudes
from Paper I.  More details about the known
members can be found there.  
\item[$^{\dagger}$] Magnitudes from this work.  When the galaxy was
not completely in our field of view, magnitudes are taken from the NASA
Extragalactic Database, and the reader is referred there for the original
sources.  When $R$ magnitudes are not available, $B$ magnitudes are
quoted.
\item[$^{*0}$] NGC 3977 is a background grand-design spiral galaxy
with a heliocentric velocity of 5722 km s$^{-1}$.
\item[$^{*1}$] NGC 3888 is a background late-type galaxy at $z=0.008$.
\item[$^{*2}$] NGC 3898 is centered just off this field to the East,
and is only partly visible.  It is a member of the nearby 12-3 Group
and has a heliocentric velocity of 1176 km s$^{-1}$. 
\item[$^{*3}$] NGC 3850 is a member of the nearby 12-3 group, with a
heliocentric velocity of 1140 km s$^{-1}$ (Verheijen et al.~2000). 
\item[$^{*4}$] UGC 6828 is a grand-design spiral, presumably background. 
\item[$^{*5}$] UGC 6956 is centered just off this field to the East,
and is only partly visible in our field of view.
\item[$^{*6}$] PC1200+4755 is a 
foreground emission-line galaxy with a spectroscopic
redshift  $z=0.002$ 
(Schneider et al.~1994).
\item[$^{*7}$] CGCG215$-$022 is a giant red early-type galaxy, presumably
background.  
\item[$^{*8}$] UGC 7069 is a luminous flat late-type galaxy not seen in HI
(Verheijen et al.~2000) and therefore presumably not a cluster member.   
\item[$^{*9}$] NGC 3811 is a background late-type galaxy at $z=0.010$.
\item[$^{*10}$] NGC 4047 is a background late-type galaxy at $z=0.011$.
\item[$^{*11}$] UGC 7358 is a background late-type galaxy at $z=0.012$.
\end{list}
\end{table*} 

\subsection{First Observing Run (1996)}

Images were taken of the fields listed in Table 1. 
All images were taken at the prime focus of the CFHT on Mauna Kea, using the
UH8K mosaic camera, a mosaic of eight 4K $\times$ 2K CCDs (Metzger 
et al.~1995; scale 0.22 arcsec pix$^{-1}$; total field of view 0.5 degree
$\times$ 0.5 degree).  The total area surveyed was then 2.2 square
degrees.  Each field was imaged as a set of three 1200 seconds 
exposures in the $R$-band, dithered by up to an arcminute 
to reject cosmic rays and bad pixels, and to ensure that regions that fell in
the gaps between the CCDs on any particular exposure were imaged in at
least one other exposure.
The $R$-band filter was chosen for this survey 
to maximize the magnitude limit for the detection of
low surface-brightness galaxies with this instrumental
setup; the unthinned
CCDs had low
quantum efficiency at shorter wavelengths and airglow emission contaminates at
longer wavelengths. 
All images were
dark-subtracted and flat-fielded using twilight sky flats.
Due to geometric
distortions arising from the large size of the camera, 
we did not combine the images until after the galaxy
detection stage since these geometric transforms
alter the noise statistics in a complex way
(see Section 3; to compensate for this effect we ran the
detection programs at a low significance threshold so as not to miss any
marginally-detected galaxies).
Instrumental magnitudes were computed from observations of standard stars,
and the photometry was converted to the Cousins $R$ magnitude system of
Landolt (1992).  Images taken under 
non-photometric conditions were calibrated initially 
using the data of Tully et al.~(1996, hereafter Paper~I) and eventually using 
the data we obtained
during the third observing run of the current program (see Section 2.4).  
The median seeing was 1.0 arcseconds.

\subsection{Second Observing Run (1999)}

Images were taken of the fields listed in Table 1. 
All images were taken at the prime focus of the CFHT on Mauna Kea, using the
CFH12K mosaic camera, a mosaic of twelve 4K $\times$ 2K CCDs (Cuillandre 
et al.~1999; scale 0.22 arcsec pix$^{-1}$; total field of view 0.7 degree
$\times$ 0.5 degree).  The total area surveyed was then 15.8
square degrees.  Each field was imaged for
420 seconds.  Subsequent exposures were progressively shifted a half--field
diameter.  Hence, most parts of the sky along the major and minor axes were
imaged twice.  The projection of camera gaps or flaws shift between exposures
so all parts of the sky along these axes were imaged at least once.  
The fields are designated by the letter A along the major axis (along with a 
number between 1 and 33 
in decreasing order of declination) and by the letter B along the minor
(along with a
number between 4 and 16 in increasing order of right ascension; fields
B01--03 and B17--22 were covered in the VLA survey but not the optical one).
Exposures were taken in the $R$-band in order to maintain consistency
with the 1996 data.  The areas covered by the combined
datasets are shown in Figure 1.   
All images were
bias-subtracted (the dark current was negligible)
and flat-fielded using twilight sky flats.
Instrumental magnitudes were computed from observations of standard stars,
and the photometry was again converted to the Cousins $R$ magnitude system of
Landolt (1992).  Some of the images were taken under 
marginally non-photometric conditions.
Originally, photometric zero-points were obtained using
the data on luminous galaxies of Paper~I with interpolation across
the overlapping fields.  However, the CFHT12K
mosaic camera is large enough that extinction due to cirrus could vary
across the detector by up to 0.2 magnitudes along the long axis of
the chip (this happened
in one or two of the images,
where the average extinction was as much as one magnitude; in most images
it was much smaller).  The entire
dataset was recalibrated once the data from the 2000 observing run was
reduced. 
The median seeing was 0.8 arcseconds.

\subsection{Third Observing Run (2000)}

Images were taken of the candidate cluster members that we identified in the
1996 and 1999 data (see Sections 3 and 4). 
All images were taken at the f/8 Cassegrain
focus of the University of Hawaii 2.2 m Telescope, also on Mauna Kea, using a
Tektronix 2048 $\times$ 2048 thinned CCD
(scale 0.22 arcsec pix$^{-1}$; field of view 7.5 arcmin 
$\times$ 7.5 arcmin).  
All data were taken under photometric conditions.
Each object was imaged for 360 seconds with a $B$ filter, 180 seconds with
an $R$ filter, and 180 seconds with an $I$ filter. 
All images were
bias-subtracted (the dark current was negligible)
and flat-fielded using twilight sky flats.
Instrumental magnitudes were computed from observations of a large number
(about 100 in each filter) of standard stars,
and the photometry was again converted to the 
Johnson ($B$) -- Cousins ($RI$) magnitude system of
Landolt (1992).  
The photometric zero-points were accurate to approximately one percent. 
The median seeing was 0.8 arcseconds.

\section{3 Photometry and membership considerations} 

In a diffuse environment like the Ursa Major Cluster, at $R>16$ the number of
background galaxies is higher than the number of cluster 
members 
(both the 1996 and 1999 datasets).  Statistical background subtraction,
as done in distant clusters (e.g.~Trentham 1998), will 
not be sufficient to allow us to correct for background contamination.
We need to take the morphologies of the galaxies that we detect into account.
As outlined in Section 1, members can be identified based on the
magnitude vs.~central surface-brightness relation of Binggeli (1994). 
For galaxies of a given apparent magnitude, cluster members that are dwarfs 
have lower surface-brightnesses and larger sizes than background giants
of the same apparent magnitude.  The light is much less concentrated
in the dwarfs.  
Our approach is to make a detailed study of the 1996 dataset
plus a number of background fields (two blank UH8K fields 
near the cluster -- the ``Blank 1'' and ``Blank 3'' fields -- 
taken in 1996 and
nine CFH12K fields taken in 1999)     
and derive a condition for membership candidature
based on the concordance between measured light concentrations (quantified
according to Binggeli's correlation)  
and the existence of a number of objects (19 in the 1996
dataset) having lower light concentrations 
than any galaxies of
equivalent apparent magnitude in 
any of the background fields. 
We then apply these criteria to the far bigger 1999 dataset.  
Note that dwarf spheroidals and dwarf irregulars follow the same 
magnitude vs.~central surface-brightness correlation
(Binggeli 1994) so that if the objects in the 1996 fields are mostly dwarf
spheroidals but those in the 1999 fields are mostly
dwarf irregulars, 
this will not bias our results.  

Our application to the 1996 data was then as follows.
Objects were detected above local sky (Poisson noise-dominated,
1$\sigma$ between 27 and 28 $R$ mag arcsec$^{-2}$)
at a low (2$\sigma$) significance level using the FOCAS detection
algorithm (Jarvis \& Tyson 1981; Valdes 1982, 1989). 
For each detected galaxy we then define an inner concentration parameter
based on aperture $R$ magnitudes:
$${\rm ICP} = R ( < 4.4\,\,{\rm arcsec}) - R ( < 2.2\,\,{\rm arcsec}),$$ 
and an outer concentration parameter: 
$${\rm OCP} = R ( < 12\,\,{\rm arcsec}) - R ( < 6\,\,{\rm arcsec}).$$
Both the ICP and OCP are more negative for galaxies of larger scale
length, which for a given apparent magnitude equates
to galaxies of lower surface-brightness.
Both are
close to zero for stars, since the seeing was always much less than
2.2 arcsec (the seeing was always good enough that its effect on
the concentration parameters for all the galaxies that we consider here
was negligible).  These concentration parameters characterize the light
distribution on physical scales between about 0.2 kpc and 1 kpc (in Ursa
Major one arcsecond is equivalent to 0.09 kpc).   
The concentration parameters for galaxies in the 1996 dataset are presented
in Figure 2. 
We now define two conditions:
$${\rm ICP} < -1.1 \eqno({\rm C1})$$ 
and
$${\rm OCP} < -0.4. \eqno({\rm C2})$$ 
Only one object in the background fields satisfies these
conditions (which we later excluded as being an object
that would mimic a candidate member had
it turned up in a cluster field, under point (v) in the list below) yet
a significant number of low surface-brightness objects in the cluster fields
do satisfy these conditions.  One would expect normal dwarf galaxies to 
satisfy both
conditions, given Figure 1 of Binggeli (1994 -- see the lines in Fig.~2).  
We therefore regard 
objects that satisfy both conditions as possible cluster members.  Figure~2 
shows that the 
differential between
members and non-members as defined by these conditions    
is not completely clean: some objects which satisfy these conditions
are certainly background objects due to their being either (i) grand-design
luminous spirals with very negative concentration parameters due to star
formation in spiral arms at large distance from the galaxy center, 
(ii) merging 
galaxies with no well-defined center, (iii) objects with a neighboring
galaxy or star that did not get separated into two objects by the detection
algorithm, (iv) extremely flat edge-on galaxies (dwarfs are relatively stubby),
or (v) low surface-brightness
material that seems to be debris or ejecta associated with a nearby giant 
galaxy. 
These cases are easily excluded from the sample.  A few objects are
less straightforward to exclude as background objects since they may show
some of these signatures at a low level; these are the objects that we
categorize ``2'' or ``3'' later in this section and discuss individually
in Section 4. 
In the entire 1996 dataset there were 4154 extended objects with a 6
arcsecond aperture magnitude $R(6) < 21.5$, only 130 of
which were left after imposing condition {\bf C1}.  Those that pass condition
{\bf C1} were mostly normal
background galaxies towards the lower end of the surface-brightness
distribution (condition {\bf C1} was a very conservative one), 
but also included clusters members, 
objects satisfying (i) -- (v) above, and some borderline cases.  
All these objects were studied by eye and the different kinds of objects
identified.   
After imposing condition {\bf C2}, there were 40 objects left, 17 of which
were excluded as being candidate members based on criteria
(i) through (v).  The remaining 23 were identified as probable or
possible members.  Four were marginal as regards satisfying these
conditions and/or have some morphological hints that they could be
background objects. 
We were less certain about these and categorize them ``2''
or ``3'' later in this section.
In addition there was one object amongst the 130 that satisfied {\bf C1}
but failed to satisfy {\bf C2} that is probably background, but in our
judgment could be an extreme cluster member; we categorize this one ``3''
as discussed later in this section.
Finally, we excluded one very flat low surface-brightness object that
appeared to move with respect to the background galaxies between exposures;
this may have been a small comet or alternatively an internal reflection
within the camera.  

\begin{figure}
\begin{center}
\vskip-2mm
\epsfig{file=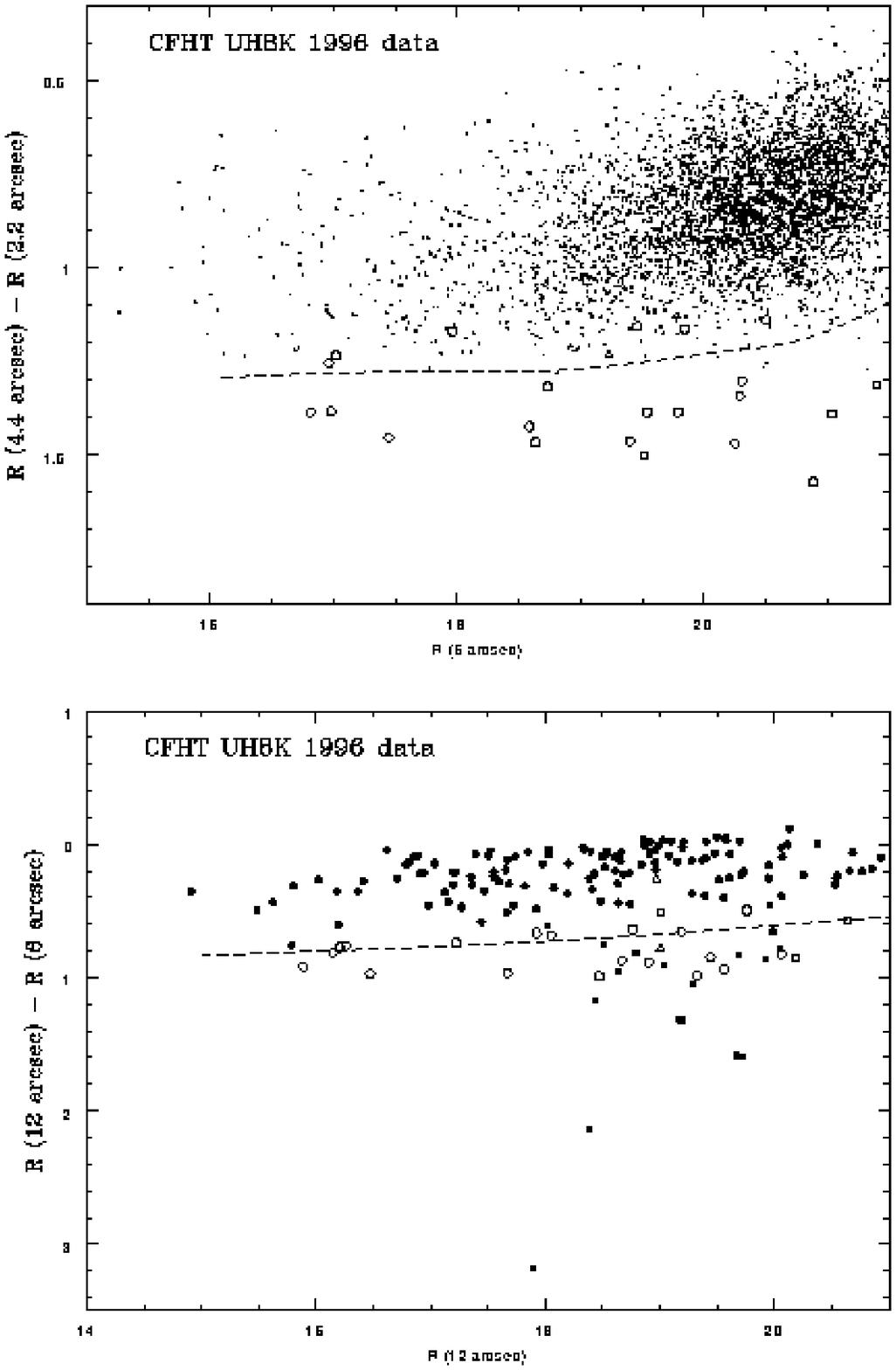, width=8.65cm}
\end{center}
\vskip-4mm
\caption{
The concentration parameters for galaxies 
in the 1996 dataset, as defined in the text, as a function of apparent
magnitude.  The top panel shows the inner concentration parameter
for all extended objects with $R(6) < 21.5$ as a function of  of apparent
$R$ magnitude measured in a 6 arcsecond circular aperture.
The dots represent objects we regard as 
as background.  The open symbols represent objects classified members or
possible members (circles = rated ``1''; squares = rated ``2'';
triangles = rated ``3'').  The dashed line is the 
median predicted position for
dwarf spheroidal galaxies, assuming the magnitude vs.~surface-brightness
correlation of Binggeli (1994), exponential light profiles (Binggeli \&
Cameron 1991), $B-R = 1.5$ (Trentham 1998 Section 5 and references     
therein), and the same observing conditions as for our 1996 data.  
The lower panel shows the outer concentration parameter
for all extended objects as a function of  of apparent
$R$ magnitude measured in a 12 arcsecond circular aperture.
Only objects in the upper panel with $R(4.4) - R(2.2) < -1.1$ are included.
Objects believe to be background are labeled as filled circles or squares,
depending on the absence or presence of a nearby object.  The open symbols
and the dashed line have the same meaning as in the upper panel.  
Background objects close to or below the 
dashed line were identified as such by
morphology (e.g.~grand-design spiral, or extreme flatness -- see points (i)
through (v) in Section 3 of the text) or by the
presence of a companion making the outer concentration parameter anomalously
negative.   
}
\end{figure}

\begin{figure}
\vskip 40mm
\caption{
The CFHT $R$-band 
images for candidate members, along with their membership ratings in brackets.
In all images north is up and east is to the left. 
The horizontal bar in each image represents 6 arcseconds.
}
\end{figure}

\begin{table*}
\caption{New members and candidates}
{$$\vbox{
\halign {\hfil #\hfil & \hfil #\hfil &
#\hfil \hfil & #\hfil \hfil && \quad \hfil \hfil #\cr

Galaxy & Rating & Comments$^{\dagger}$ & Field$^{*}$ &  
$\alpha$ (2000) & $\delta$ (2000) 
& ICP & OCP & $R_{\rm TOT}$ & $M_R$ & $B-R$ & $R-I$ & aperture&\cr 

umd  &   & &   &  & &  & & & & & & arcsec   &\cr  
\noalign{\smallskip} \noalign{\smallskip}
\cr 
\noalign{\smallskip}

01 & 2 & AB & B4 &
$11 \, 36 \, 04.7 \, $&
$47 \, 31 \, 14 \, $&
$-1.21$ & $-0.50$ & 
17.69 
& $-$13.66 
& $1.00 \pm 0.06$ & $0.32 \pm 0.08$   & 6 &\cr

02 & 3 & A & B5 &
$11 \, 39 \, 27.5 \, $&
$47 \, 34 \, 10 \, $&
$-1.18$ & $-0.68$ &
17.77 
& $-$13.58
& $0.80 \pm 0.07$ & $0.08 \pm 0.12$   & 6 &\cr

03 & 2 & & B6  &
$11 \, 44 \, 10.9 \, $&
$48 \, 02 \, 24 \, $&   
$-1.00$ & $-0.63$ &
19.13 
& $-$12.22
& $0.84 \pm 0.15$ & $0.23 \pm 0.21$   & 6 &\cr

04 & 1 & & A2 &
$11 \, 45 \, 16.3 \, $&
$55 \, 34 \, 31 \, $&
$-1.64$ & $-0.50$ &
18.84 
& $-$12.51
& $1.18 \pm 0.34$ & $0.28 \pm 0.41$   & 6 &\cr

05 & 3  & C &  B6 &
$11 \, 45 \, 57.9 \, $&
$47 \, 37 \, 16 \, $&
$-1.13$ & $-1.12$ &
18.84 
& $-$12.51
& $0.70 \pm 0.12$ & $0.28 \pm 0.22$   & 6 &\cr

06 & 3 & C & A1 &  
$11 \, 46 \, 19.0 \, $&
$56 \, 02 \, 17 \, $&
$-1.07$ & $-0.40$ &
16.67 
& $-$14.68
& $1.05 \pm 0.02$ & $0.31 \pm 0.03$   & 6 &\cr

07 & 0 & & A2 &  
$11 \, 46 \, 35.0 \, $&
$55 \, 49 \,  16 \, $&
$-1.24$ & $-0.69$ &
16.12 
& $-$15.23
& $0.84 \pm 0.02$ & $0.29 \pm 0.03$   & 6 &  \cr

08 & 3 & D & A4 &
$11 \, 46 \, 35.9 \, $&
$54 \, 47 \, 55 \, $&
$-1.24$ & $-0.72$ &
18.53 
& $-$12.82
& $1.43 \pm 0.21$ & $0.42 \pm 0.22$   & 6 & \cr

09 & 0 & & B7  &
$11 \, 46 \, 49.6 \, $&
$48 \,   05 \,  33 \, $&
$-1.14$ & $-0.56$ &
17.80 
& $-$13.55
& $0.91 \pm 0.07$ & $0.28 \pm 0.03$   & 6 & \cr

10 & 1 & & A4 &
$11 \, 46 \, 53.4 \, $&
$54 \, 40 \, 10 \, $&
$-1.77$ & $-0.76$  &
18.50 
& $-$12.85
& $1.22 \pm 0.18$ & $0.25 \pm 0.22$   & 6 & \cr

11 & 1 & & A4 &
$11 \, 47 \, 13.9 \, $&
$54 \, 35 \, 57 \, $&
$-1.25$ & $-0.80$ &
18.14 
& $-$13.21
& $1.22 \pm 0.30$ & $0.76 \pm 0.28$   & 6 & \cr

12 & 1 & E & A3 &
$11 \, 47 \, 22.3 \, $&
$55 \, 26 \, 10 \, $&
$-1.36$ & $-0.76$ &
17.65 
& $-$13.70
& $1.31 \pm 0.21$ & $0.45 \pm 0.23$   & 6 &\cr

13 & 1 & E & A3 &
$11 \, 47 \, 33.5 \, $&
$55 \, 11 \, 03 \, $&
$-1.21$ & $-0.49$ &
17.62 
& $-$13.73
& $1.26 \pm 0.06$ & $0.38 \pm 0.07$   & 6 & \cr

14 & 2 & A & A3 &
$11 \, 48 \, 12.6 \, $&
$55 \, 10 \, 26 \, $&
$-1.35$ & $-0.55$ &
18.07 
& $-$13.28
& $0.43 \pm 0.11$ & $0.18 \pm 0.19$   & 6 &\cr

15 & 2 & A & A2 &
$11 \, 48 \, 43.6 \, $&
$55 \, 55 \, 43 \, $&
$-1.28$ & $-0.77$ &
16.56 
& $-$14.79
& $1.09 \pm 0.06$ & $0.19 \pm 0.08$   & 6 & \cr

16 & 1 & E & A1 &
$11 \, 48 \, 45.4 \, $&
$56 \, 01 \, 56 \, $&
$-1.23$ & $-0.51$ &
19.15 
& $-$12.20
& $1.26 \pm 0.22$ & $0.29 \pm 0.26$   & 6 &\cr

17 & 3 & & A4 &
$11 \, 49 \, 18.7 \, $&
$54 \, 58 \, 15 \, $&
$-0.94$ & $-0.42$ &
19.77 
& $-$11.58
& $0.60 \pm 0.19$ & $0.02 \pm 0.33$   & 6 &\cr

18 & 2 & E &A3 &
$11 \, 49 \, 26.3 \, $&
$55 \, 15 \, 13 \, $&
$-1.40$ & $-1.20$ &
17.29 
& $-$14.06
& $0.96 \pm 0.27$ & $0.98 \pm 0.26$   & 6 & \cr

19 & 1 & E & A3 &
$11 \, 50 \, 45.6 \, $&
$55 \, 06 \, 45 \, $&
$-1.20$ & $-0.56$ &
18.58
& $-$12.77
& $1.32 \pm 0.16$ & $-0.14 \pm 0.24$   & 6 & \cr

20 & 2 &  & A4 &
$11 \, 50 \, 50.7 \, $&
$54 \, 46 \, 00 \, $&
$-1.05$ & $-0.71$ &
19.96 
& $-$11.39
& $1.59 \pm 0.51$ & $0.13 \pm 0.55$   & 6 & \cr

21 & 0 & & A8 &
$11 \, 51 \, 53.6 \, $&
$53 \,  05 \,  59 \, $&
$-1.20$ & $-0.66$ &
15.75 
& $-$15.60
& $0.72 \pm 0.01$ & $0.31 \pm 0.01$   & 6 & \cr

22 & 1 & EF & N3953, A10 &
$11 \, 53 \, 09.2 \, $&
$52 \, 11 \, 22 \, $&
$-1.45$ & $-0.97$ &
15.27 
& $-$16.08
& $0.99 \pm 0.02$ & $0.41 \pm 0.03$   & 6 & \cr

23 & 0 & & B8 &
$11 \, 53 \, 11.1 \, $&
$48 \,  11 \, 18 \, $&
$-1.23$ & $-0.61$ &
17.46 
& $-$13.89
& $0.59 \pm 0.05$ & $0.44 \pm 0.08$   & 6 & \cr

24 & 3 & C & A11 &
$11 \, 53 \, 52.3 \, $&
$51 \, 29 \, 38 \, $&
$-1.14$ & $-0.56$ &
16.67 
& $-$14.68
& $0.61 \pm 0.02$ & $0.20 \pm 0.04$   & 6 & \cr

25 & 1 & AEG & A12 &
$11 \, 54 \, 27.6 \, $&
$51 \, 20 \, 05 \, $&
$-1.26$ & $-0.66$ &
17.68 
& $-$13.67
& $1.01 \pm 0.10$ & $0.22 \pm 0.14$   & 6 &\cr

26 & 3 & B & B9 &
$11 \, 54 \, 40.8 \, $&
$48 \, 13 \, 49  \, $&
$-1.11$ & $-0.34$ &
19.45 
& $-$11.90
& $1.05 \pm 0.18$ & $-0.40 \pm 0.35$  & 6 &\cr

27 & 1 & H & N3998 &
$11 \, 55 \, 38.2 \, $&
$55 \,  22 \,  04 \, $&
$-1.57$ & $-0.82$ &
19.45 
& $-$11.90
& $0.13 \pm 0.44$ & $0.83 \pm 0.55$   & 6 & \cr

28 & 3 & B & U6917 & 
$11 \, 55 \, 53.3 \, $&
$50 \,  31 \,  12 \, $&
$-1.23$ & $-0.26$ &
18.90 
& $-$12.45
& $0.75 \pm 0.08$ & $0.16 \pm 0.12$   & 6 & \cr  

29 & 2 & & B9 &
$11 \, 55 \, 59.1 \, $&
$48 \, 12 \, 02 \, $&
$-1.03$ & $-0.58$ &
18.85 
& $-$12.50
& $1.46 \pm 0.19$ & $-0.21 \pm 0.28$   & 6 &\cr

30 & 1 & E & N3998 &
$11 \, 56 \, 09.4 \, $&
$55 \,  15 \,  54 \, $&
$-1.46$ & $-0.64$ &
18.13 
& $-$13.22
& $1.41 \pm 0.16$ & $0.48 \pm 0.16$   & 6 &\cr

31 & 1 & E & N3998 &
$11 \, 57 \, 01.6 \, $&
$55 \,  25 \,  10 \, $&
$-1.23$ & $-0.76$ &
15.66 
& $-$15.69
& $1.17 \pm 0.01$ & $0.37 \pm 0.01$   & 18 &\cr

32 & 1 & E & N3998 &
$11 \, 57 \, 03.1 \, $&
$55 \,  25 \,  12 \, $&
$-1.39$ & $-0.78$ &
15.75 
& $-$15.60
& $1.21 \pm 0.01$ & $0.39 \pm 0.01$   & 18 &\cr

33 & 1 & E & N3992 &
$11 \, 57 \, 03.8 \, $&
$53 \,  18 \,  03 \, $&
$-1.47$ & $-0.49$ &
19.10 
& $-$12.25
& $1.24 \pm 0.16$ & $0.00 \pm 0.23$   & 6 &\cr

34 & 1 & E & N3992 &
$11 \, 57 \, 05.6 \, $&
$53 \,  26 \,  28 \, $&
$-1.16$ & $-0.66$ &
17.72 
& $-$13.63
& $1.27 \pm 0.19$ & $0.26 \pm 0.22$   & 6 &\cr

35 & 1 & E & N3992 &
$11 \, 57 \, 21.0 \, $&
$53 \,  13 \,  35 \, $&
$-1.42$ & $-0.67$ &
17.22 
& $-$14.13
& $1.05 \pm 0.03$ & $0.34 \pm 0.04$   & 12 &\cr

36 & 1 & I & N3992 &
$11 \, 57 \, 36.6 \, $&
$53 \,  10 \,  01 \, $&
$-1.34$ & $-0.85$ &
18.71 
& $-$12.64
& $0.53 \pm 0.13$ & $0.15 \pm 0.21$   & 12 &\cr

37 & 1 & E & N3998 &
$11 \, 58 \, 02.7 \, $&
$55 \,  14 \,  48 \, $&
$-1.16$ & $-0.99$ &
18.08 
& $-$13.27
& $1.24 \pm 0.11$ & $0.41 \pm 0.13$   & 12 &\cr

38 & 0 & & A17 &
$11 \, 58 \, 11.6 \, $&
$48 \, 52 \,  55 \, $&
$-1.41$ & $-1.06$ &
14.76 
& $-$16.59
& $0.81 \pm 0.01$ & $0.29 \pm 0.01$   & 24 &\cr

39 & 1 & E & N3998 &
$11 \, 58 \, 13.7 \, $&
$55 \,  23 \,  16 \, $&
$-1.25$ & $-0.81$ &
15.69 
& $-$15.66
& $1.56 \pm 0.01$ & $0.48 \pm 0.01$   & 24 &\cr

40 & 0 & & A17 &
$11 \, 58 \, 26.0 \, $&
$48 \,  57 \,  36 \, $&
$-1.36$ & $-0.71$ &
17.68 
& $-$13.67
& $0.58 \pm 0.16$ & $0.15 \pm 0.25$   & 6 &\cr

41 & 1 & E & N3992 &
$11 \, 58 \, 34.3 \, $&
$53 \,  20 \,  44 \, $&
$-1.17$ & $-0.74$ &
16.79 
& $-$14.56
& $1.27 \pm 0.02$ & $0.44 \pm 0.02$   & 12 &\cr

42 & 2 & & N3992 &
$11 \, 58 \, 47.9 \, $&
$53 \,  27 \,  14 \, $&
$-1.31$ & $-0.57$ &
20.58 
& $-$10.77
& $1.36 \pm 0.49$ & $0.35 \pm 0.52$   & 6 &\cr

43 & 0 & & A16 &
$11 \, 59 \, 57.6 \, $&
$49 \,  33 \,  50 \, $&
$-1.36$ & $-0.76$ &
14.43 
& $-$16.92
& $0.98 \pm 0.01$ & $0.40 \pm 0.01$   & 24 &\cr

44 & 0 & & A20 &
$12 \,  00 \, 35.3 \, $&
$47 \, 46 \, 24 \, $&
$-1.48$ & $-1.08$ &
13.70 
& $-$17.65
& $0.85 \pm 0.03$ & $0.38 \pm 0.04$   & 6 &\cr

45 & 3 & C & A19 &
$12 \, 00 \, 35.6 \, $&
$47 \, 58 \, 10 \, $&
$-1.00$ & $-0.37$ &
19.58 
& $-$11.77
& $1.04 \pm 0.15$ & $0.13 \pm 0.21$   & 6 &\cr

46 & 2 & C & A21 &
$12 \, 02 \, 21.3 \, $&
$47 \, 07 \, 36 \, $&
$-1.13$ & $-0.44$ &
17.40 
& $-$13.95
& $1.19 \pm 0.04$ & $0.36 \pm 0.04$   & 6 &\cr

47 & 0 & & A25 &
$12 \, 02 \, 43.7 \, $&
$45 \,  11 \,  28 \, $&
$-1.16$ & $-0.72$ &
14.55 
& $-$16.80
& $1.25 \pm 0.01$ & $0.47 \pm 0.01$  & 24 &\cr

48 & 1 &   & A29 &
$12 \, 04 \, 10.3 \, $&
$43 \, 39 \, 17 \, $&
$-1.15$ & $-0.53$ &
19.35 
& $-$12.00
& $1.17 \pm 0.23$ & $-0.32 \pm 0.42$   & 6 &\cr

49 & 1 & I & A27 &
$12 \, 04 \, 49.9 \, $&
$44 \, 26 \, 34 \, $&
$-1.28$ & $-0.64$ &
18.28 
& $-$13.07
& $1.07 \pm 0.17$ & $-0.44 \pm 0.34$   & 6 &\cr

50 & 1 & E & A28 &
$12 \, 05 \, 24.8 \, $&
$43 \, 42 \, 32 \, $&
$-1.46$ & $-1.04$ &
15.17 
& $-$16.18
& $1.18 \pm 0.05$ & $0.39 \pm 0.05$   & 12 &\cr

51 & 1 & & N4100 &
$12 \, 05 \, 45.5 \, $&
$49 \,  42 \,  54 \, $&
$-1.30$ & $-0.98$ &
18.83 
& $-$12.52
& $1.13 \pm 0.31$ & $0.17 \pm 0.41$   & 6 &\cr

52 & 1 & E & N4100 &
$12 \, 06 \, 05.2 \, $&
$49 \,  28 \,  47 \, $&
$-1.14$ & $-0.94$ &
19.76 
& $-$11.59
& $1.28 \pm 0.38$ & $-1.82 \pm 1.43$   & 6 &\cr

53 & 1 & E & N4100 &
$12 \, 06 \, 05.8 \, $&
$49 \,  25 \,  37 \, $&
$-1.32$ & $-0.68$ &
17.75 
& $-$13.60
& $1.27 \pm 0.06$ & $0.36 \pm 0.07$   & 6 &\cr

54 & 2 & & A30 &
$12 \, 06 \, 26.0 \, $&
$42 \, 54 \, 33 \, $&
$-1.13$ & $-0.44$ &
18.67 
& $-$12.68
& $1.15 \pm 0.18$ & $0.18 \pm 0.24$   & 6 &\cr

55 & 2 & & N4100 & 
$12 \, 06 \, 26.7 \, $&
$49 \,  33 \, 24 \, $&
$-1.39$ & $-0.85$ &
20.56 
& $-$10.79
& $0.74 \pm 0.45$ & $0.09 \pm 0.68$   & 6 &\cr

56 & 1 & & A30 &
$12 \, 06 \, 27.9 \, $&
$43 \, 01 \, 14 \, $&
$-1.13$ & $-0.27$ &
19.48 
& $-$11.87
& $1.06 \pm 0.26$ & $0.43 \pm 0.31$   & 6 &\cr

57 & 1 & & A28 &
$12 \, 07 \, 09.4 \, $&
$43 \, 59 \, 15 \, $&
$-1.14$ & $-0.43$ &
19.61 
& $-$11.74
& $1.21 \pm 0.32$ & $0.16 \pm 0.41$   & 6 &\cr

58 & 2 & I & B12 &
$12 \, 07 \, 44.7 \, $&
$48 \, 51 \, 23 \, $&
$-1.36$ & $-0.61$ &
19.80 
& $-$11.55
& $0.87 \pm 0.21$ & $0.03 \pm 0.34$   & 6 &\cr

59 & 2 & AB & U7176 &
$12 \, 09 \, 28.1 \, $&
$50 \,  12 \,  42 \, $&
$-1.50$ & $-0.51$ &
19.58 
& $-$11.77
& $0.86 \pm 0.15$ & $0.23 \pm 0.21$   & 6 &\cr

60 & 1 & G & U7176 &
$12 \, 09 \, 34.4 \, $&
$50 \,  26 \,  02 \, $&
$-1.46$ & $-0.96$ &
17.19 
& $-$14.16
& $0.79 \pm 0.08$ & $0.20 \pm 0.12$   & 6 &\cr

 &  & &   &    &   & &             &            &           &   &    &\cr
           \noalign{\smallskip}
\noalign{\smallskip}\cr}}$$}
\end{table*}

\begin{table*}
{$$\vbox{
\halign {\hfil #\hfil & \hfil #\hfil &
#\hfil \hfil & #\hfil \hfil && \quad \hfil \hfil #\cr

Galaxy & Rating & Comments$^{\dagger}$ & Field$^{*}$ &
$\alpha$ (2000) & $\delta$ (2000)
& ICP & OCP & $R_{\rm TOT}$ & $M_R$ & $B-R$ & $R-I$ & aperture&\cr

umd  &   & &   &  & &  & & & & & & arcsec   &\cr
\noalign{\smallskip} \noalign{\smallskip}
\cr
\noalign{\smallskip}

61 & 3 & J & U7176 & 
$12 \, 09 \, 47.8 \, $&
$50 \,  03 \,  51 \, $&
$-1.13$ & $-0.78$ &
19.34 
& $-$12.01
& $0.55 \pm 0.15$ & $-0.10 \pm 0.29$   & 6 &\cr

62 & 1 & & U7176 &
$12 \, 11 \, 13.3 \, $&
$50 \, 06 \, 41 \, $&
$-1.39$ & $-0.87$ &
18.71 
& $-$12.64
& $1.10 \pm 0.19$ & $0.09 \pm 0.27$   & 6 &\cr

63 & 1 & EK & U7176 &
$12 \, 11 \, 22.7 \, $&
$50 \, 16 \, 11 \, $&
$-1.39$ & $-0.92$ &
15.11 
& $-$16.24
& $1.23 \pm 0.01$ & $0.40 \pm 0.01$   & 24 &\cr

64 & 1 & E & U7176 &
$12 \, 11 \, 34.9 \, $&
$50 \, 26 \, 13 \, $&
$-1.39$ & $-0.89$ &
18.92 
& $-$12.43
& $1.37 \pm 0.26$ & $0.03 \pm 0.34$   & 6 &\cr

65 & 2 & & B14 &
$12 \, 14 \, 43.7 \, $&
$49 \, 01 \, 28 \, $&
$-1.01$ & $-0.41$ &
20.28 
& $-$11.07
& $0.94 \pm 0.52$ & $-0.18 \pm 0.84$   & 6 &\cr
 &  & &   &    &     & &           &                        &   &    &\cr
           \noalign{\smallskip} 
\noalign{\smallskip}\cr}}$$}
\begin{list}{} {}
\item[$^{*}$] See Table 1 for the designations.  
\vskip 1pt \noindent
For the 1996 data, the 
fields are named for luminous giant galaxies.  
\vskip 1pt \noindent
For the 1999 data, the fields
are labelled according
to their position along the major or minor axis. 
\vskip 7pt
\item[$^{\dagger}$] Legend for the third column:
\vskip 1pt
A: No HI detection, although clearly irregular/late-type morphology  
\vskip 1pt
B: weak evidence for a bar  
\vskip 1pt
C: high central surface-brightness (i.e.~small physical size) if a cluster
dwarf 
\vskip 1pt
D: possibly associated with a nearby galaxy  
\vskip 1pt
E: smooth morphology; probably a dSph  
\vskip 1pt
F: huge extended halo  
\vskip 1pt
G: blue colour   
\vskip 1pt
H: extreme low surface-brightness 
\vskip 1pt
I: irregular morphology; probably a dIrr 
\vskip 1pt
J: extremely flat; probably background 
\vskip 1pt
K: two nucleii; probably a recent merger 
\vskip 1pt
\end{list}
\end{table*}

The prescription outlined in the previous paragraph can now be applied to
the 1999 dataset, which is much larger (there were about 30,000
galaxies with $R(6) < 21.5$). 
All objects satisfying {\bf C1} and all that were within 0.1 magnitudes of
satisfying both {\bf C1} and {\bf C2} were looked at individually.
Contaminant objects according to the criteria (i) through (v) above were 
excluded
upon inspection by eye. 
Additionally we found a number of bright late-type galaxies that 
satisfied  both {\bf C1} and {\bf C2} but were not
seen in HI, and questioned their membership on these grounds (see Section 4
for further details).   Most of the difficult cases are found at the north end 
of the major axis, close to the 12-3 Group (Tully 1987$a$), 
and could arise from contamination
by that group.  Note that the HI 
observations did not cover the entire velocity range of the
12--3 group.

It will be clear from the previous two paragraphs that we are somewhat more
confident about the membership possibilities of some objects than others.
We therefore introduce the following subjective
rating scheme, based on our
own assessment, for all objects we find other than those presented in Paper~I. 
Candidates are characterized ``0'' to ``3'', where
\vskip 1pt \noindent
``0'': membership confirmed from HI data (Verheijen et al.~2000,
Verheijen et al.~2001).  There
were no new HI detections in the 1996 fields, but nine new detections in the
1999 fields; 
\vskip 1pt \noindent
``1'': probable member, but no HI detection; 
\vskip 1pt \noindent
``2'': possibly a member, but conceivably background; 
\vskip 1pt \noindent
``3'': probably background, but conceivably a member.   
\vskip 1pt \noindent
Our judgments are based primarily on by how clearly each object satisfies
{\bf C1} and {\bf C2}, the morphological criteria (i)
through (v), and on the implications of
no HI detection in bright
galaxies with distinctly irregular morphologies.  The
justifications for our rating of each object are listed in the next section, 
along with
comments on some objects that we excluded as being possible candidates.

For each candidate we estimate its total $R$ magnitude as follows.
(i) We compute the
flux within some isophote slightly above the extraction limit set by
the sky background.
This isophote corresponded to the 1.8$\sigma$ isophote in the 1996 data
and the 1.5$\sigma$ isophote in the (less deep) 1999 data.
For a few extremely large low surface-brightness galaxies, 
the galaxies could reliably be followed beyond this isophote to a fainter one.
(ii) We compute the amount of light
that falls below the specified
isophote at large radius by 
fitting an exponential light profile to a part of the galaxy where there
are no condensations and extrapolating this light profile beyond the
last fitted isophote to infinity (Paper~I).  The results are 
estimates of the total, as opposed
to isophotal magnitudes.  The uncertainties in these magnitudes depend
on the details of the light distribution beyond the isophotal radius
for each galaxies, which are uncertain, but are likely to be far less than
the bin size (2 mag) that we shall use in computing luminosity functions.  
We only consider galaxies whose isophotal magnitudes are brighter than 
$R=21.5$.  At fainter magnitudes it is difficult to rate
with confidence any galaxies 
as ``1'' since
low surface-brightness galaxies in the background fields begin to appear at 
these faint limits.  The faintest magnitude for our luminosity
function is therefore not determined by detection constraints, but by
where we lose the ability to distinguish cluster members from background
galaxies on surface-brightness grounds.  
Apparent magnitudes were then converted to absolute magnitudes assuming
a distance of 18.6 Mpc, corresponding to a distance modulus of 31.35.
Colours (from the 2000 data) were computed from apertures centered on the
$R$-band galaxy center.
Using aperture magnitudes ensures that we are probing the same stellar
populations in all filters.  The aperture sizes were between 6 and 24
arcseconds.  Larger apertures were used for bigger galaxies so as to improve
the signal-to-noise.  Errors in the colours are dominated  
by Poisson sky noise and are large for the faintest galaxies.  
The apertures were
always large enough so that differential seeing effects between the different
filters were always negligible.

Clearly our methods of identifying members {\it a priori}
bias us towards selecting a particular kind of galaxy --
normal dwarf galaxies. 
By far most local dwarfs are normal dwarf spheroidals or dwarf
irregulars (Binggeli 1994), but it is 
still instructive to see how we fare with other kinds of low luminosity 
galaxies like blue compact dwarfs (BCDs).
Most BCDs have irregular morphologies 
(Telles, Terlevich \& Melnick 1997) and extended star-formation
and/or low surface-brightnesses.  We expect these objects to satisfy both
{\bf C1} and {\bf C2} and so to appear in our samples.
As a test, we placed the well-studied BCD galaxy UGC 6456 (Lynds et al.~1998)
in the Ursa
Major Cluster and were able to recover it using the above strategy. 
In addition, we expect BCDs to be detected in HI and so turn up in the
VLA sample.

There are, however, two kinds of objects that we {\it do} miss, given our
selection criteria.
\vskip 1pt \noindent
(i) We do not find extreme low surface-brightness disks, which have
central surface-brightnesses below 27 $R$ mag arcsec$^{-2}$.  No such
galaxies are known (although they would be extremely 
difficult to find anywhere;
e.g.~Disney 1976).  If they did exist and were gas-rich we would find them
in our HI survey.
The disks of Malin 1 (Bothun et al.~1987),
F568$-$6 (Bothun et al.~1990) and GP1444 (Davies,
Phillipps \& Disney 1988), would have
easily turned up in our HI survey, and probably in our optical survey as
well.   
It is only gas-poor extreme low surface-brightness galaxies we need to
worry about (the stars in such hypothetical objects
could be gravitationally bound by a dark matter halo,
for example). 
\vskip 1pt \noindent
(ii) We also miss galaxies with smooth de Vaucouleurs light profiles but
faint ($> 20$ mag arcsec$^{-2}$) central surface-brightnesses, because they
look like background ellipticals.  Recall from the fundamental plane (see
Kormendy \& Djorgkvoski 1989) 
that higher luminosity ellipticals have fainter central  
surface-brightnesses, so that if we observe an elliptical galaxy
with a moderate central surface-brightness, it is {\it a priori}  
likely to be a background luminous
galaxy. Galaxies with the surface brightness profiles of
elliptical galaxies but that lie so far away from the fundamental plane
as to cause misidentifications with the background appear to be
rare, but there was at least one
in our survey.  Markarian 1460, a Blue Compact Galaxy, failed to satisfy 
{\bf C1}
and {\bf C2} very substantially, but is a known cluster member
based on optical spectroscopy (Pustilnik et al.~1999) and on an HI
detection (Verheijen et al.~2001). 
Only this one object is known in the cluster (Trentham, Tully \& Verheijen 
2001), although cases with less HI could be missed.

\section{Sample}

As described in the previous section, we now have a list of possible members, 
each with
a rating ``0'' to ``3'' depending on our confidence regarding membership. 
These are listed in Table 2, along with the coordinates
and photometric properties.  
Images from the CFHT data are presented in Figure 3.  
The objects are numbered in order of increasing right ascension. 
For the galaxies we detect we make various specific
comments in the third column of Table 2; these often indicate why
a particular galaxy received a particular rating.

We do not attempt classifications as dwarf spheroidals or irregulars 
(although see the list below for more obvious cases) 
because such classifications are not reliable based on morphological
information alone (eg, some ellipsoidal, textureless dwarfs have HI and some 
lumpy dwarfs are not detected in HI). 

Below are comments on some of the more interesting objects
that satisfied the above
selection criteria but we excluded from our sample.  Where no names exist, 
we derive them from the J2000 coordinates: hhmm.m(+/-)ddmm.  
These are:
\vskip 1pt
\noindent
{\bf NGC 3850} -- This galaxy easily satisfied both {\bf C1} and {\bf C2},
but its heliocentric velocity of 1140 km s$^{-1}$ (Verheijen et al.~2001)
places it in the nearby 12-3 Group and not in Ursa Major.  We therefore
exclude it from the sample;   
\vskip 1pt
\noindent
{\bf 1145.9+5605} -- This very bright ($R \sim 14$) late-type galaxy would
surely have been seen in HI were it in the cluster, but it was not.  It
therefore must be a background object.  Additionally we see signs of
weak spiral structure, which would be consistent with this being a
luminous background late-type galaxy;
\vskip 1pt
\noindent
{\bf 1154.7+5053} -- This late-type peculiar galaxy ($R \sim 17.5$) shows weak 
spiral structure,
including a number of knots embedded in one arm, which are probably
HII regions.  This object was not seen in HI and so is presumably background; 
\vskip 1pt
\noindent
{\bf 1154.2+5053} -- This bright ($R \sim 16$) late-type peculiar galaxy 
was not seen in HI and so is presumably background;  
\vskip 1pt
\noindent
{\bf PC1200+4755} -- This irregular low surface-brightness galaxy
easily satisfied both {\bf C1} and {\bf C2}, but it is a foreground
object given its spectroscopic redshift
$z=0.002$  (Schneider, Schmidt \& Gunn 1994);
\vskip 1pt
\noindent
{\bf 1152.9+4754} -- This bright ($R \sim 17.5$) late-type peculiar galaxy
was not seen in HI and so is presumably background.  It has two nucleii
and weak spiral structure, which would support this interpretation;  
\vskip 1pt
\noindent
{\bf 1155.5+4846} -- This bright ($R \sim 15.5$) interacting peculiar galaxy  
was not seen in HI and so is presumably background;  
\vskip 1pt
\noindent
{\bf 1207.1+4259} -- This bright ($R \sim 16$) late-type peculiar galaxy
was not seen in HI and so is presumably background.  It has 
some spiral structure with many condensations (presumably HII regions)
embedded in the arms, which would support this interpretation;
\vskip 1pt
\noindent
{\bf 1206.4+4226} -- This bright ($R \sim 16.5$) flattened late-type galaxy 
was not seen in HI and so is presumably background;
\vskip 1pt
\noindent
{\bf 1209.6+4201} -- This bright ($R \sim 14.5$) late-type galaxy
was not seen in HI and so is presumably background.  There is weak
evidence for spiral structure, which would support this interpretation;
\vskip 1pt
\noindent
{\bf 1213.4+4901} -- This bright ($R \sim 16.5$) late-type galaxy
was not seen in HI and so is presumably background.  Again there is weak
evidence for spiral structure, which would support this interpretation.
\vskip 1pt
\noindent
Additional notes on these and other galaxies that were excluded are presented
in the notes to Table 1.
 
\section{Luminosity functions}

The luminosity functions are presented in Figure 4.  It is immediately
apparent that wherever we set the borderline between members and
non-members (based on our ratings), the Ursa Major luminosity function
is much shallower than the Virgo one. 
Values of $\alpha$ were computed from power-law fits to the data and the
results are presented in Table 3.  The value of $\alpha$ 
appropriate to Ursa Major is about $-1.1$, with an uncertainty of
about 0.2.  In comparison, in the Virgo Cluster 
 $\alpha = -2.26 \pm 0.14$ from
a similar power-law fit over the range $-16 < M_R < -11.5$ (Phillipps et 
al.~1998).  Note, however, that in this study of the Virgo Cluster
there was no culling of the sample for background objects, 
as in the present paper, which might lead to a slight 
overestimate of the Virgo numbers.  

\begin{figure}
\begin{center}
\vskip-2mm
\epsfig{file=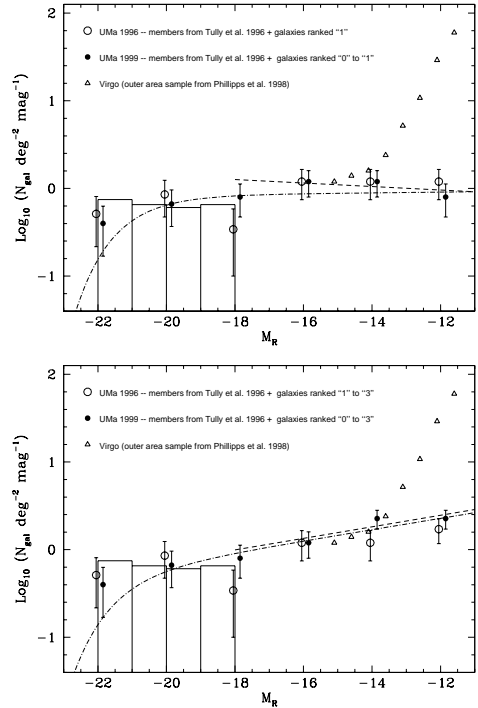, width=8.65cm}
\end{center}
\vskip-4mm
\caption{The luminosity function of the Ursa Major Cluster.  
The 1996 data is offset horizontally
by 0.2 mag to permit a clearer comparison with the 1999 data.
The Virgo Cluster data by Phillipps et al.~(1998)
and the 1996 data are normalized to have the same number of $M_R = -15$
galaxies rated ``0'' or ``1'' as in the 1999 data.  
The histograms represent the luminosity functions for the 
bright-galaxy sample of
Papers I and II, assuming a completeness limit of $M_R = -18$ and
a normalization set by scaling the total number of galaxies with $M_R < -18$
to be the same as for the 1999 dataset.
The dashed lines represent the best-fitting power-laws to the 1999
data (the filled circles)
for $M_R > -18$; these have $\alpha = -0.95$ (upper panel) and
$\alpha = -1.16$ (lower panel). 
The dotted-dashed lines represent the best-fitting Schechter (1976) functions
to the joint Paper I + II ($M_R < -18$) and 1999 ($M_R > -18$) 
datasets.  
These fits have
($M_R^*,\alpha^*)$ = $(-21.44,-1.01)$ (upper panel) and
$(-21.71,-1.16)$ (lower panel).
}
\end{figure}

\begin{table*}
\caption{Power law fits to lumiosity functions}
{$$\vbox{
\halign {\hfil #\hfil && \quad \hfil #\hfil \cr

                & Galaxies rated ``1'' &  
Galaxies rated ``1'' -- ``2'' & Galaxies rated ``1'' -- ``3'' &\cr 
\noalign{\smallskip} \noalign{\smallskip}
\cr 
\noalign{\smallskip}
1996 data & & & &\cr
$-16.85 < M_R < -10.85$ & $\alpha = -1.00 \pm 0.18$ 
& $\alpha = -1.04 \pm 0.17$ & 
$\alpha = -1.10 \pm 0.16$ &\cr
$-18.85 < M_R < -10.85$ & $\alpha = -1.08 \pm 0.16$ & 
$\alpha = -1.10 \pm 0.15$ & 
$\alpha = -1.16 \pm 0.14$ &\cr
 &   &     &       &\cr 
1999 data & & & &\cr
$-16.85 < M_R < -10.85$ & $\alpha = -0.90 \pm 0.18$
& $\alpha = -1.07 \pm 0.14$ &
$\alpha = -1.15 \pm 0.13$ &\cr
$-18.85 < M_R < -10.85$ & $\alpha = -1.00 \pm 0.12$ &
$\alpha = -1.12 \pm 0.10$ &
$\alpha = -1.18 \pm 0.09$ &\cr
 &   &     &       &\cr   
                   \noalign{\smallskip} 
\noalign{\smallskip}\cr}}$$}
\begin{list}{}{}
\item[]Known members from Tully et al.~(1996) are included in
all the fits, in addition
to the new candidates.  
\end{list}
\end{table*}

The Virgo and Ursa Major luminosity functions are therefore highly
inconsistent with each other.  There is one caveat: the
Ursa Major sample may be incomplete if there exist substantial numbers
of either (see Section 3)
gas-poor extreme low-surface-brightness galaxies or
low-luminosity galaxies that look like distant elliptical or S0 galaxies
due to the combination of central surface-brightnesses around 19 $R$
mag arcsec$^{-2}$ and smooth de Vaucouleurs light profiles (like
Markarian 1460). 
In order for the Ursa Major luminosity function to be the same as the
Virgo one, we would need to have missed many hundred such objects.
Objects of the first type do not 
appear to be common: on going to fainter magnitudes
we did not find proportionately more galaxies with extremely low surface
brightnesses. 
The lowest surface brightness object
that we detected was umd 27, which had an average surface
brightness of 24.6 $R$ mag arcsec$^{-2}$ within the 20\% light radius,
somewhat brighter than the 1-$\sigma$ limit quoted in Section 3.
Given that the only case of the second
type known is Markarian 1460, we do not
regard as a serious worry either.  
 
Since the Ursa Major Cluster is a diffuse unevolved cluster of spiral
galaxies and since most galaxies in the Universe reside in such environments,
our results may be representative of the field luminosity function
as well, down to $M_R = -11$ ($M_B \sim -10$). 
The Local Group is another example of such an environment and has a
similarly shallow luminosity function ($\alpha = -1.1$; van den Bergh 1992),
albeit with large errors due to poor counting statistics.
More negative values of $\alpha$ (i.e.~steeper luminosity functions)
have been found in the spectroscopic
surveys of Lin et al.~(1996) and Marzke et al.~(1994), although these
values of $\alpha$ come from a Schechter (1976) function fit to the
bright galaxies, not a power-law fit to galaxies as faint as the ones
we observe here.  
The values of $\alpha$ we present are only valid for the magnitude ranges 
identified in
Table 3, which only overlap marginally at the bright end with the
spectroscopic samples.  Given this fact, and the difference in the way
$\alpha$ is computed from the data, the two results are not 
inconsistent.  

The fraction of HI--detectable objects is decreasing toward
fainter absolute magnitudes.  Nevertheless some of the faintest optical
galaxies (like umd 40) do have significant amounts of cold gas.  A
more quantitative investigation of this phenomenon is presented in
Verheijen et al.~(2001).

\section{Colours}

The colours of the new galaxies are listed in Table 2 and presented in 
Figures 5 and 6.  In Figure 7 we present a histogram of the $B-R$ colours
of galaxies whose errors in $B-R$ are less than 0.2 mag.  
The galaxies we consider here all have $-18 < M_R < -10$ if they lie
in Ursa Major, hence qualify as dwarfs (see Binggeli 1994).  

\begin{figure}
\begin{center}
\vskip-2mm
\epsfig{file=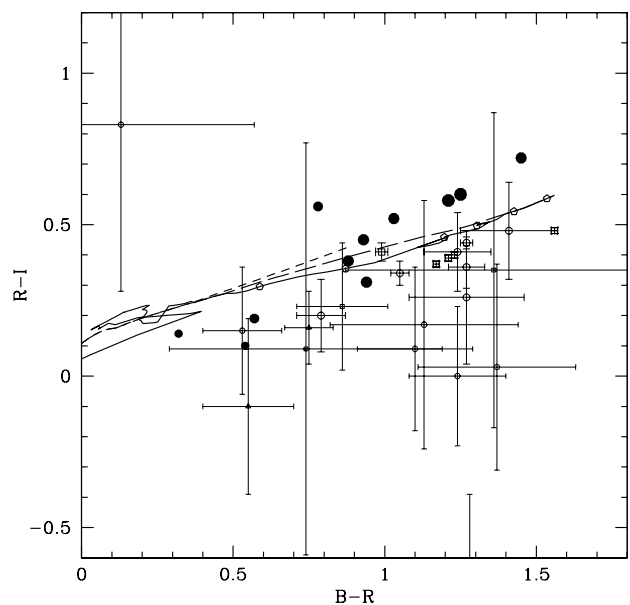, width=8.65cm}
\end{center}
\vskip-4mm
\caption{
Colour-colour diagram for the 1996 data, with colours
computed as described in the text. 
The symbols have the following meanings: filled circles -- confirmed members
(from Paper~I); open circles -- galaxies rated ``1'';
open squares -- galaxies rated ``2'';
open triangles -- galaxies rated ``3''.
Larger symbols represent galaxies of higher luminosity.
The lines are the stellar population evolutionary tracks from the models of
Bruzual \& Charlot (1993), computed assuming a metallicity of $Z=0.008$
(0.4 solar)
and a Salpeter (1995) initial mass function with mass limits of 0.1 and 100
M$_{\odot}$.  
The different lines have the following meanings:
solid line -- instantaneous burst;
short dashed line -- continuous star formation model;
long dashed line -- exponentially decaying star formation model with $e$-folding
timescale of 1 Gyr.
The open pentagons on the instantaneous burst
model lines represent the colours at 0.5, 1, 2,
4, 8, and 16 Gyr after the burst, in order of increasing
$B-R$.  
The error bars represent Poisson errors from sky subtractions. 
}
\end{figure}

\begin{figure}
\begin{center}
\vskip-2mm
\epsfig{file=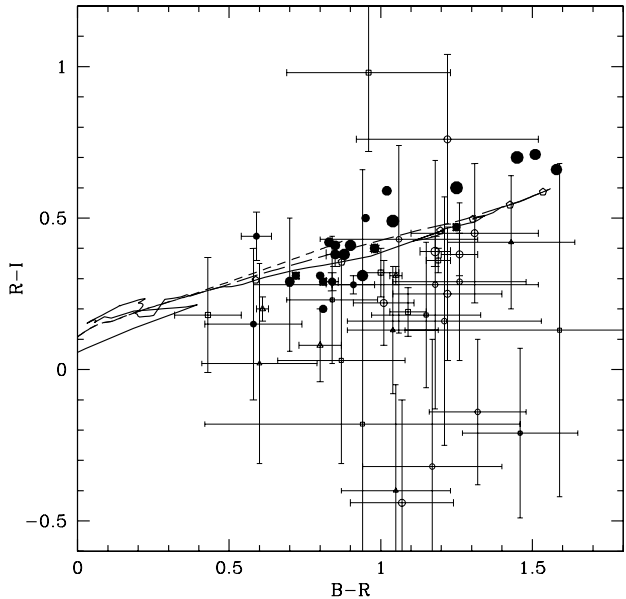, width=8.65cm}
\end{center}
\vskip-4mm
\caption{
Colour-colour diagram for the 1999 data.  The lines and
symbols have the same meanings as in Figure 5, except that the filled
circles now also represent galaxies rated ``0''.
}
\end{figure}

\begin{figure}
\begin{center}
\vskip-2mm
\epsfig{file=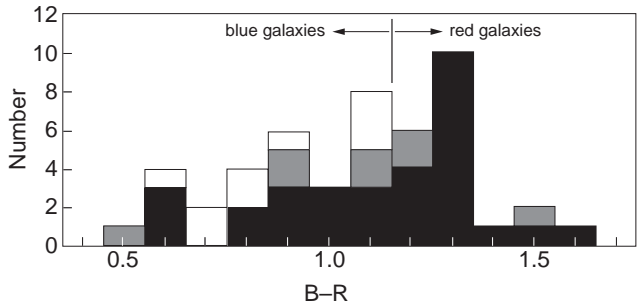, width=8.65cm}
\end{center}
\vskip-4mm
\caption{
Histogram of 
$B-R$ colours of the new galaxies.  Both the 1996
and 1999 data are plotted.  Only galaxies with an error of $< 0.2$ mag in
$B-R$ are included.  The solid histogram represents galaxies rated ``0'' or 
``1'', the shaded portion represents galaxies rated ``2'', and the open 
portion, galaxies rated ``3''.  The division between ``blue'' and ``red''
dwarfs at $B-R = 1.15$, as described in the text, is shown. 
}
\end{figure}

Typical values are $1.2 < B-R < 1.6$ for dSph galaxies and 
$B-R < 1.1$ for dIrr galaxies (see Coleman, Wu \& Weedman 1980 and 
Section 5 of Trentham
1998; see also references within those papers). 
With $B-R = 1.15$ as a division between ``red'' and ``blue'' dwarfs,
we find that in the 1996 data there are 10 (12) red dwarfs rated ``0'' or
``1'' and 5 (7) blue dwarfs (where the number before the brackets is 
restricted to cases with colour errors $<0.2$ and the number in brackets 
includes cases with large colour errors).  In the 1999 data 
there are 5 (11) red dwarfs rated ``0'' and  
``1'' and 10 (12) blue dwarfs.
Most of the 
red dwarfs have featureless morphologies, indicating that they are indeed
dSph galaxies. 
The fraction of dwarfs that are dSph was
higher in the 1996 dataset than in the 1999 dataset, 
suggesting that the
dSphs might cluster around luminous galaxies more than dIrrs {\it within}
the Ursa Major Cluster, just as Binggeli
et al.~(1990) found in the local Universe.   

By contrast with the Ursa Major Cluster, in the Virgo Cluster
the vast majority
of low luminosity galaxies are dwarf spheroidals (Sandage et al.~1985,
Phillipps et al.~1998).

\section{Distribution of dwarfs in the cluster}

The total area covered by the 1999 dataset was about 7.7 times the 
area covered by the 1996 dataset,  yet we only found 1.2 times as many
galaxies rated ``0'' or ``1''.  It therefore appears that galaxies are
somewhat more likely to be found around luminous galaxies in the cluster than
in random places in the cluster.   
This tendency is more marked for red dSph galaxies than for blue dIrr ones.

\begin{figure}
\begin{center}
\vskip-2mm
\epsfig{file=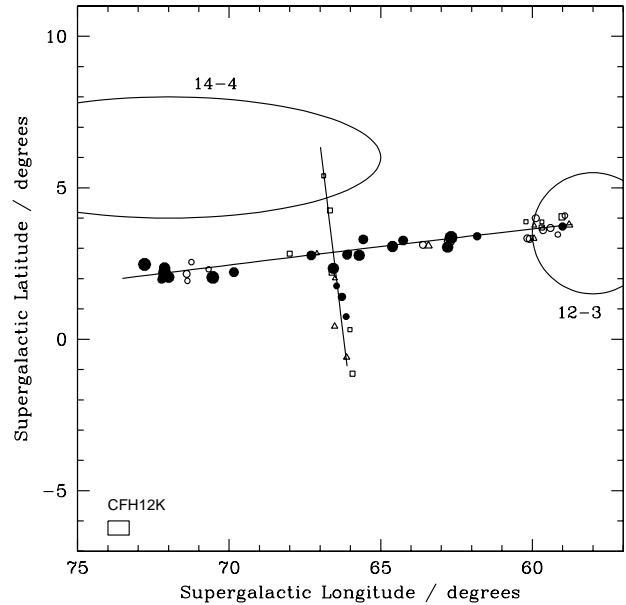, width=8.65cm}
\end{center}
\vskip-4mm
\caption{
Projected positions of confirmed and candidate members
in supergalactic coordinates.  
The symbols have the following meanings: filled circles -- confirmed members
(from Paper~I and galaxies
rated ``0''); open circles -- galaxies rated ``1'';
open squares -- galaxies rated ``2'';
open triangles -- galaxies rated ``3''.
Larger symbols represent galaxies of higher luminosity.
The lines are the locus of the field centers; the CFH12K field size is shown by
the box in the lower right corner.  The locations of the nearby  
12$-$3 and 14$-$4 groups are shown. 
}
\end{figure}

There is a strong suggestion of a
correlation between the tendency to have a
substantial satellite population and parent luminosity 
within the 1996 dataset.
The four galaxies with the largest satellite populations are 
NGC 3992 (Sbc; $M_R=-21.93$; 6 probable/possible dwarfs), 
NGC 3998 (E/S0; $M_R=-21.80$; 6 probable dwarfs), 
NGC 4157 (Sb; $M_R=-21.34$; 5 probable/possible dwarfs), and
NGC 4100 (Sbc; $M_R=-21.24$; 4 probable/possible dwarfs).
Note that NGC 4157 was just to the north of
the field centered on UGC 7176.
There were four luminous Sb-bc galaxies in the 1996
dataset, only one not manifesting a significant satellite population
(NGC 3953: Sbc; $M_R=-21.86$; 1 probable dwarf).  
The three cluster fields observed in 1996 centered on a low luminosity
galaxy or no known member were completely devoid of new dwarf candidates. 

One of the two 1996 fields with the largest number
of dwarfs (six) contains the NGC 3998/3990/3972 subgroup, an aggregation of 
three
luminous galaxies (including two S0s), which may be a region of the cluster
with an anomalously low local crossing time.  
One of these galaxies, NGC 3998, is the only galaxy in Ursa Major seen to have
a large globular cluster population. 

The 1996 fields were chosen based on prior information on the positions of
galaxies so an unbiased analysis
of the clustering properties within these fields is not possible (they were
chosen because prior WSRT observations of the fields were available).
On the other hand, the 1999 fields were chosen to simply march along the
major and minor axes in the cluster,
so they represent a reasonable random sampling.  In Figure~8, we show the 
positions of the galaxies in the 1999 fields in supergalactic coordinates.  
It is clear from this 
figure that our sampling intersects two aggregations of dwarfs, one at each 
end of the major axis.  

The aggregation at the high supergalactic longitude end lies in the general
vicinity of the subcondensation of early-type galaxies,
NGC 4111, 4117, and 4143, in and adjacent field A30.
The only other subcondensation of early type galaxies in the Ursa Major 
Cluster, that around NGC 3998, was found from the 1996 data to be an
environment elevated in dwarfs.

The aggregation at the low supergalactic longitude end might suffer
serious contamination from the nearby 12-3 Group.
The 12-3 Group is nearby in the sky to this end of the major axis of
the cluster but at slightly higher velocity than the Ursa Major Cluster.
In Paper~I it is noted
that it is at this border with the 12-3 Group
that the Ursa Major Cluster is least cleanly defined.  It is distinctly
possible that the dwarfs in this region identified as possible/probable group 
members are {\it not} in the Ursa Major Cluster but, rather, are in the 
adjacent 12-3 Group and it will take velocity information or precise
distance discrimination to settle the question.  We are assuming that these
galaxies {\it are} members of the Ursa Major Cluster, so if they are not
members the luminosity function has even a lower slope than we claim.

Outside these two condensations, the clustering of dwarfs within the
1999 dataset is weak.  The principal finding is the {\it paucity} of
dwarf candidates.  Along the major and minor axis strips, it is 
as likely to find a big, previously cataloged cluster member as a 
dwarf galaxy 
probably/possibly associated with the cluster.  There are so few candidate
dwarfs that it makes little sense to look for spatial correlations.

We have found that the red spheroidal candidates do tend to reside near
most of the luminous giant galaxies in our survey region, though they
are not abundant even there.  Elsewhere they are rare.  The blue candidates,
some with detectable HI and others with only HI upper limits, are more
loosely distributed through the cluster but in remarkably small numbers.

\section{Discussion and Summary}

The following eight observations and 
inferences present a summary view of aspects of the Ursa Major
Cluster (see also Paper~I).  
The last six derive from the present work.  For each of these new
results we
discuss possible reasons for why the cluster behaves the way it does in the
context of the wider galaxy formation problem.
\vskip 1pt \noindent
(1) The Ursa Major Cluster is a
diffuse 
cluster of 62 luminous galaxies with $M_B<-16.9$ (assuming a distance 
$d=18.6$~Mpc), all but eight of which are
late-type galaxies (Paper~I).  
The early-type galaxies in the cluster are 
lenticulars, with at most one elliptical, and none are extremely luminous.  
Five of
the early-type galaxies lie in two nests at opposite ends of the cluster.  
\vskip 1pt \noindent
(2) The mass of the Ursa Major Cluster is $5 \times 10^{13} {\rm M}_{\odot}$,
about one-twentieth the mass of the Virgo Cluster (the $B$-band mass-to-light
ratio is 6 times higher for Virgo than for Ursa Major; see the
introduction to Paper I) .
The collapse timescale for the Ursa Major Cluster is $\sim 17$~Gyrs,
roughly a Hubble time (Tully~1987$b$).
By comparison, the collapse timescale for the Virgo Cluster is
$\sim 2$~Gyrs, so that cluster is much more dynamically evolved.
Most galaxies in the Universe exist in unevolved environments like the Ursa
Major Cluster. 
This cluster is therefore a
good place to study dwarf galaxy properties in an environment akin to the 
field yet the density of galaxies is high enough that counting statistics have
significance.
\vskip 1pt \noindent
(3) The luminosity function of dwarf galaxies in the Ursa Major Cluster 
is very flat, with a logarithmic slope $\alpha = -1.1$ and an uncertainty
not more than 0.2.  This distribution is far shallower than the Virgo Cluster 
luminosity
function, which has $\alpha = -2.26 \pm 0.14$ (Phillipps et al.~1998).
Dwarf galaxies are thought to be heavily dark-matter dominated 
(e.g.~Aaronson \& Olszewski 1987,
Persic \& Salucci 1988,  Kormendy 1990,  Pryor \& Kormendy 1990).  
Cold dark matter hierarchical clustering theories of galaxy 
formation predict large numbers of these small
dark-matter halos everywhere, a robust prediction that follows directly
from the fluctuation spectrum.  
It appears that the expectations of the theory
are met, and {\it only} met, in dynamically evolved regions like the
Virgo Cluster. 
Hence there is apparently {\it suppression} in low density
environments.

Galaxies within evolved clusters have had many crossing times in which to
interact.
Gas cannot continue to be accreted onto galaxies
once they enter cluster environments since the gas is outside the Roche-limit
of the galaxies within the cluster potential (Shaya \& Tully 1984).  
That gas thermalizes and
becomes the X-ray plasma.  The environment in low density regions is
much more benign for gas accretion onto the dwarfs.  {\it In spite of this} 
the dwarfs are numerous in dynamically evolved environments and not in
low density environments.  

Suppose that dwarfs that have ended up in evolved clusters formed very early,
on top of the large scale fluctuation that grew into the cluster.  Most of
them have been absorbed into bigger systems but there were such large numbers 
of them
that many survive.  Over in the low density regions the entire process
was slowed.  The small scale perturbations that produced dwarfs collapsed
over a more extended time.  Supernovae and massive star formation
periodically evacuate or rarefy the gas.  If nothing else was going on,
perhaps this gas could re-accumulate and continue to form stars.  But
something else {\it is} going on: 
in the low density phase the gas is being ionized
by the metagalactic ultraviolet flux.  
The dark wells might persist but gas cannot 
accumulate to form stars (Klypin et al.~1999, Bullock et al.~2000).  
The result is a dwarf luminosity function that is steep in clusters but
flat in the field.
In such a scenario, the dwarf formation in Virgo happens much earlier
than the dwarf formation in Ursa Major; this timescale difference is
consistent with what is expected given hierarchical cluster models
of galaxy formation given a plausible redshift for the reionization of
the Universe (Tully et al.~2001). 
\vskip 1pt \noindent
(4) Dwarfs in the Ursa Major Cluster are as likely to be dwarf irregulars as
dwarf spheroidals.  This finding derives particularly from the 1999 dataset, 
where
we observed blind fields along the minor and major axes.
Many low luminosity galaxies therefore appear to be capable of ongoing
star formation, as is the case with the giant galaxies in the cluster.  
This property is
probably linked to the young dynamic age of the cluster.
The galaxies in question have been relatively isolated all their lives and 
have been
able to draw upon gas reservoirs over an extended time.
The Ursa Major Cluster is quite
different from the Virgo Cluster, where HI gas reservoirs have more frequently
been depleted
and the dwarf galaxies that we see there today are mostly dwarf spheroidals. 
\vskip 1pt \noindent
(5) Most dwarfs that we detect have reasonably low surface brightnesses 
and follow the absolute magnitude vs. surface brightness correlation
suggested by Figure 1 of Binggeli (1994; also see Tully \& Verheijen 1997, 
Paper~II). 
Compact dwarfs are rare.
The same is true in both the local field (Binggeli, Sandage \& Tammann 1988)
and in richer clusters like Coma (Karachentsev et al.~1995)
where the luminosity function
of high surface brightness ellipticals is falling rapidly at faint magnitudes. 
\vskip 1pt \noindent
(6) The dwarf galaxies seem to be clustered within the Ursa Major Cluster. 
In particular, there appear to be two aggregations, each of about twelve
galaxies, at each end of the major axis.
Dwarfs probably have substantial dark-matter halos 
which themselves are highly clustered in the context of
hierarchical clustering models
of galaxy formation.
Clustering {\it within} the Ursa Major Cluster 
has not yet been washed out by galaxy-galaxy interactions. 
This circumstance is not surprising given the low velocity dispersion, hence 
long crossing time.
\vskip 1pt \noindent
(7) All but one of the giant galaxies targetted in the 1996 observations 
have a significant number of spheroidal candidates projected within $150$~kpc.
Perhaps their presence manifests a small-scale version of the physical 
processes discussed in point (3) since the {\it local} dynamical time
on the mass scale of the giant galaxy is short.  Alternatively, since
the numbers of dwarfs are modest, these objects could be debris from
ancient galaxy formation interactions (Mirabel, Dottori, \& Lutz, 1992)  
\vskip 1pt \noindent
(8) Galaxies with HI detections can be as faint as $R=19$ ($M_R = -12$). 
The details of the
HI observations are 
presented elsewhere (Verheijen et al.~2000). 

\section*{Acknowledgments}

NT acknowledges the PPARC for financial support.
Helpful discussions with M.~Disney, R.~Terlevich, and E.~Terlevich are
gratefully acknowledged. 
Observations were made at the Canada-France-Hawaii Telescope, which is
operated by the National Research Council of Canada, the Centre
National de la Recherche Scientifique de France, and the University of
Hawaii.

\end{document}